\documentclass[fleqn,usenatbib]{mnras}

\usepackage{newtxtext,newtxmath}

\usepackage[T1]{fontenc}
\usepackage{ae,aecompl}

\usepackage{graphicx}	
\usepackage{amsmath}	
\usepackage{xcolor}

\usepackage{etoolbox}


\newcommand{\poisson}{{\sl Poisson}}

\newcommand{\alos}{{\sl $A_{\rm los}$}}

\newcommand{\galah}{{\sl GALAH}}

\newcommand{\segue}{{\sl SEGUE}}
\newcommand{\sdss}{{\sl SDSS}}
\newcommand{\apogee}{{\sl APOGEE}}

\newcommand{\rave}{{\sl RAVE}}
\newcommand{\lamost}{{\sl LAMOST}}

\newcommand\gaia{\textit{Gaia}}

\newcommand\gdrtwo{\gaia~DR2}
\newcommand\gedrthree{\gaia~EDR3}
\newcommand\gdrthree{\gaia~DR3}
\newcommand{\tgas}{{\gaia-TGAS}}

\newcommand{\ruwe}{\textit{RUWE}}
\newcommand{\astromexnoise}{\textit{astrometric\_excess\_noise }}
\newcommand{\visibperiods}{\textit{visibility\_periods\_used }}
\newcommand{\errorovparallax}{$\sigma_{\varpi}/\varpi$}
\newcommand{\rc}{{\sl RC}}

\newcommand{\xrvs}{{\sl xRVS}}

\newcommand{\midslice}{{\sl midslice}}
\newcommand{\upperslice}{{\sl upperslice}}
\newcommand{\topslice}{{\sl topslice}}

\newcommand{\khannalos}{{K18}}

\def\vphi{\ifmmode{\>V_{\mathrm{\phi}}}\else{$V_{\mathrm{\phi}}$}\fi}
\def\vR{\ifmmode{\>V_{R}}\else{$V_{R}$}\fi}
\def\vz{\ifmmode{\>V_{z}}\else{$V_{z}$}\fi}
\def\vlos{\ifmmode{\>V_{\mathrm{los}}}\else{$V_{\mathrm{los}}$}\fi}

\newcommand{\vphimod}{$V_{\phi,{\rm mod}}$}
\newcommand{\vRmod}{$V_{R,{\rm mod}}$}
\newcommand{\vzmod}{$V_{z,{\rm mod}}$}
\newcommand{\vall}{$V_{\phi,\rm R,z,\rm los}$}

\newcommand{\rad}{$R$}
\newcommand{\mnz}{$|z|$}

\newcommand{\mnvphi}{$\langle V_{\rm \phi} \rangle$}

\newcommand{\Rsun}{$R_{\odot}$}

\newcommand{\kms}{{\rm km s}$^{-1}$}

\newcommand{\tikzmark}[1]{\tikz[overlay, remember picture] \coordinate (#1);}
\usepackage{tikz}

\title[Streaming motion in the Milky Way disc with \gedrthree{}+]{Measuring the Streaming motion in the Milky Way disc with \gedrthree{}+.}

\author[Shourya Khanna et al.]{
Shourya Khanna$^{1,2,4}$\thanks{E-mail: shourya.khanna@inaf.it},
Sanjib Sharma$^{3,4,5}$,
Joss Bland-Hawthorn$^{3,4,6}$,
\newauthor Michael Hayden$^{3,4}$
\\
$^{1}$Kapteyn Astronomical Institute, University of Groningen, Groningen, 9700 AV, The Netherlands\\
$^{2}$INAF - Osservatorio Astrofisico di Torino, via Osservatorio 20, 10025 Pino Torinese (TO), Italy\\
$^{3}$Sydney Institute for Astronomy, School of Physics, A28, The University of Sydney, NSW, 2006, Australia\\
$^{4}$ARC Centre of Excellence for All Sky Astrophysics in Three Dimensions (ASTRO-3D)\\
$^{5}$Space Telescope Science Institute, Baltimore, USA\\
$^{6}$Miller Professor, Miller Institute, UC Berkeley, Berkeley CA 94720\\
}

\date{Accepted 2023 January 13. Received 2023 January 02; in original form 2022 May 05}

\pubyear{2023}
\begin{document}
\label{firstpage}
\pagerange{\pageref{firstpage}--\pageref{lastpage}}
\maketitle

\begin{abstract}
We map the 3D kinematics of the Galactic disc out to 3.5 kpc from the Sun, and within 0.75 kpc from the midplane of the Galaxy. To this end, we combine high quality astrometry from \gedrthree{}, with heliocentric line-of-sight velocities from \gdrtwo{}, and spectroscopic surveys including \apogee{}, \galah{}, and \lamost{}. We construct an axisymmetric model for the mean velocity field, and subtract this on a star-by-star basis to obtain the residual velocity field in the Galactocentric components (\vphi{}, \vR, \vz), and \vlos{}. The velocity residuals are quantified using the power spectrum, and we find that the peak power ($A/$[\rm \kms{}]) in the midplane ($|z|<0.25$ kpc) is ($A_{\phi},A_{\rm R},A_{\rm Z},A_{\rm los}$)=($4.2,8.5,2.6,4.6$), at $0.25 < |z|/[{\rm kpc}] < 0.5$, is ($A_{\phi},A_{\rm R},A_{\rm Z},A_{\rm los}$)=($4.0,7.9,3.6,5.3$), and at $0.5 < |z|/[{\rm kpc}] < 0.75$, is ($A_{\phi},A_{\rm R},A_{\rm Z},A_{\rm los}$)=($1.9,6.9,5.2,6.4$). Our results provide a sophisticated measurement of the streaming motion in the disc and in the individual components. We find that streaming is most significant in \vR, and at all heights ($|Z|$) probed, but is also non-negligible in other components. Additionally, we find that patterns in velocity field overlap spatially with models for Spiral arms in the Galaxy. Our simulations show that phase-mixing of disrupting spiral arms can generate such residuals in the velocity field, where the radial component is dominant, just as in real data. We also find that with time evolution both the amplitude and physical scale of the residual motion decrease.
\end{abstract}





\section{Introduction}
\label{sec:introduction}

Mapping the spatial and kinematic properties of the Milky Way disc has been an ongoing endeavour for several decades. These efforts, spread across wavelength, and sky coverage, have shown us that the Galactic disc is a complex structure. Non-axisymmetric features of varying scalelengths and scaleheights, such as the Galactic bar \citep{Babusiaux:2005,Cabrera-Lavers:2008,Wegg:2015}, and Spiral arms \citep{Reid:2004,Reid:2019,Poggio:2021}, pervade the stellar disc. Additionally, observations have shown that the disc is roughly flat out to the Solar radius, and then bends away from the plane, to give a warped appearance \citep{Drimmel:2001,Yusifov:2004,Chen:2019}. In recent years, the kinematic signatures of this stellar warp in vertical velocities, have also been mapped \citep{Poggio:2017,GaiaCollaboration:2018b,Poggio:2018}.

Much of the mapping of the disc has relied on large photometric and spectroscopic surveys, such as \sdss{} \citep{Juric:2008,York:2000}, \rave{} \citep{Steinmetz:2020}, \apogee{} \citep[][]{Majewski:2017, Jonsson:2020} , \lamost{} \citep{Cui:2012, Zhao:2012}, and \galah{} \citep{DeSilva:2015}. Such magnitude limited surveys have provided large statistical samples of stars with line-of-sight velocity (\vlos{}), and chemical abundances. Using standard-candle like tracers (ex: red clump giants, \rc{} hereafter), to estimate distances, the kinematics could then be mapped out to several kpc from the Sun. Using Red Clump stars from the RAVE survey, \cite{Williams:2013} showed that stars in the disc are participating in bulk motion, and that there are differences in the bulk motion (or streaming) North and South of the Galactic plane. In particular, for the Galactocentric radial velocity (\vR{}), they found evidence of a large outward flow above the plane, and inward flow below the plane. For the Galactocentric vertical velocity (\vz{}), they find a wave-like pattern, where, stars interior to the Solar circle and above the plane are moving upwards, while those below, downwards. Similar wave-like compression/rarefaction was also seen in both number density and bulk velocity in the \sdss{} data by \citet{Widrow:2012}, and also towards the Galactic anti-center with \lamost{} data \citep{Carlin:2013}.

The results from these studies have hinted at large scale velocity flows and fluctuations in the Galactic disc. If the Galaxy is axisymmetric, and in dynamical equilibrium, we expect negligible fluctuations in the residual velocity field. \citet[][B15 hereafter]{Bovy:2015} studied the deviations from an axisymmetric model for the line-of-sight velocity field, for a sample of red clump giants from the \apogee{} survey \citep{Bovy:2014}. They found that the power spectrum of the velocity residuals had a peak of about 11 \kms{}, and the corresponding physical scale was about 2.5 kpc. Their result suggested the presence of streaming motion on scales an order of magnitude larger than the Solar neighbourhood. Furthermore, B15 found that the peak power in the velocity residuals, could be minimised if the azimuthal component of the Solar peculiar velocity was $V_{\odot}=22.5$ \kms{}, i.e., about 10 \kms{} higher than the widely used $V_{\odot}=12.24$ \kms{} \citep[Local Standard of Rest, ][]{Schonrich:2010}. Thus, such large-scale streaming motion has important implications for the local standard of rest.

However, while the \rc{} is a very useful distance tracer, intrinsic population variance \citep{Girardi:2016,Nataf:2016} can introduce systematic errors in the distance estimates. In particular, in \citet[][K18 hereafter]{khanna1}, we showed that standard spectro-photometric schemes for selecting \rc{} stars, contaminate the samples with a high mass tail of core Helium burning stars, that are not standard-candles. The distance estimates for these high mass (brighter) contaminants are under-estimated. In K18, we expanded on the work by B15, and showed that some of the high velocity residuals in the red clump \vlos{} maps, could be due to incorrect distance estimates for the high mass tail stars. Using a \rc{} sample by combining the \galah{} and \apogee{} surveys, we probed the non-axisymmetric motion in the midplane of the Galactic disc, as well as out to 1 kpc away from the plane. As in B15, we subtracted an axisymmetric model for the \vlos{}, but allowing for flexibility in the circular velocity profile (radial and vertical), as well as on the dispersion scale length. We showed that, after taking into account, various systematics, the peak power in the velocity residuals was no more than about \alos $=6$ \kms{} in the midplane, and consistent with \poisson{} noise away from the plane.

With the advent of the \gaia{} \citep{GaiaCollaboration:2016} astrometric datasets, we now have at our disposal proper motion measurements for over a billion stars in the Galaxy. The second data release \citep[][\gdrtwo{}]{GaiaCollaboration:2018a} also provided line-of-sight velocities for about 7 million stars mostly with magnitude $G<13$ \citep[][RVS henceforth]{RVS}. In December 2020, the astrometry and photometry were updated as part of The Gaia Early Data Release 3 \citep[][\gedrthree{}]{GaiaCollaboration:2021release}. The release is based on 33 months of observations and marks a significant improvement over the previous \gdrtwo{}, bettering precision in proper motion by nearly a factor of two, and in parallax by $30\%$. Combined with radial velocities from the RVS set, this forms a very powerful dataset to probe the disc kinematics. The typical precision of proper motion catalogues in the pre-\textit{Gaia} era, of the order of 2 mas $yr^{-1}$ down to magnitude $15$ \citep{Zacharias:2017}, has now been improved by nearly two orders of magnitude. The immense impact of the improved dataset can be seen across several works that have, and keep discovering new substructure across the Galactic disc \citep{dr2kinmap2018,Antoja2018,Ramos:2018,Fragkoudi:2019,Bland-Hawthorn:2019,khanna2,Trick:2019,Hunt:2019,Monari:2019,Trick:2021,Laporte:2020,Eilers2020,gaia_anticenter2021}.

In this paper, we exploit the high precision astrometry from \gedrthree{}, and combine it with radial velocities from \gdrtwo{}, and the major spectroscopic surveys available. We use this dataset to probe the kinematics of the disc in 3-dimensions, i.e., in the individual Galactocentric velocity components as well as in the line-of-sight component. We subtract an axisymmetric model for each component and analyse the velocity fluctuations and the power spectrum in each. We compare our findings to a simulation of disrupting spiral arms, and offer a few possible scenarios for the patterns observed in the data.

\section{Datasets:}
\label{sec:datasets}

\subsection{Observational: Astrometry \& Radial velocities}

\begin{figure*}
\includegraphics[width=2.\columnwidth]{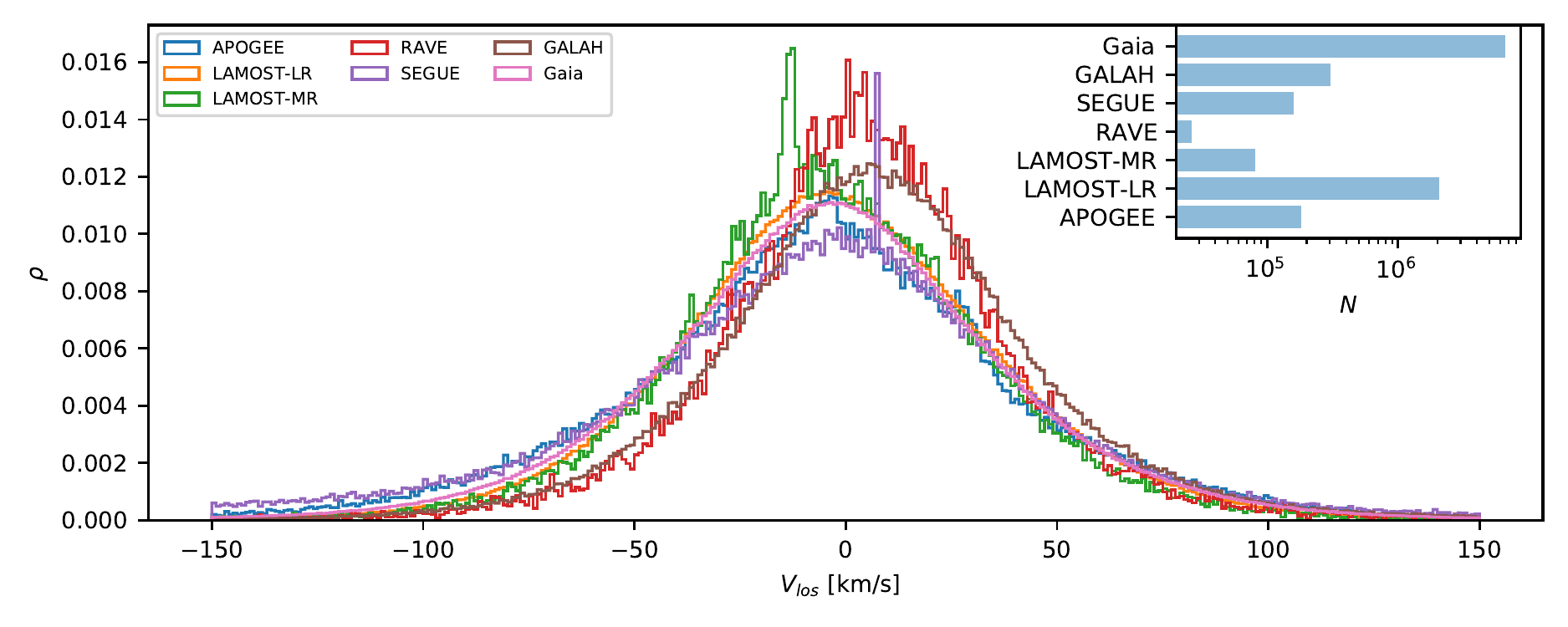} 
\caption{Line-of-sight velocities (\vlos{}) compiled by combining \gaia{} and the major spectroscopic surveys to form the \xrvs{} dataset. We show a comparison between the normalised distribution of \vlos{} from the individual surveys. The inset shows that most of the radial velocities are from \gdrtwo{}, followed by \lamost{}, and then the remaining surveys. \label{fig:xvrs_dist}}
\end{figure*}

Fortunately, in addition to \gaia{} RVS, we also have access to several ongoing spectroscopic surveys with publicly available radial velocities. These supplementary measurements vary in their sky coverage, magnitude and also precision, but in combination with the RVS, allow one to create an extended radial velocity sample (\xrvs) with high precision astrometry from \gaia{}. In particular, we crossmatch within a radius of 5 arcsec, the \gedrthree{} catalogue with data from, the \lamost{} DR6 Low Resolution \citep[LR,][]{Wang:2020}, and Medium resolution \citep[MR,][]{Liu:2019} surveys, \rave{} DR6 \citep{Steinmetz:2020}, \galah{} DR3 \citep{Buder:2021}, \apogee{} DR16 \citep{Ahumada:2020}, and \segue{} DR10 \citep{Ahn:2014}. On top of this, as recommended by the \lamost{} DR6 release note\footnote{\url{http://dr6.lamost.org/v2/doc/release-note}}, we apply a $+7.9$ \kms{} offset to the \lamost{} 
LR velocities. For those stars, whose \gdrtwo{} RVS is unavailable, we assign radial velocities in the following order: \galah{}, \apogee{}, \rave{}, \lamost{}, and finally, \segue{}. This is in accordance with the typical accuracy of the line-of-sight velocity by these spectroscopic surveys. Our initial \xrvs{} consists of 10,828,676 stars in all, and the distribution by survey is shown in \autoref{fig:xvrs_dist}. Our dataset is dominated by radial velocities from \gdrtwo{}, followed by \lamost{} $LR$, and then the remaining surveys. The typical uncertainty in the radial velocities ($\sigma_{V_{\rm los}}$) from \lamost{} $LR$, tends to be around 7 \kms{} \citep{Li:2021}, which is much higher than that in the other surveys. Thus, we restrict our sample to $\sigma_{V_{\rm los}} < 7$ \kms{}.

\begin{figure}
\includegraphics[width=1.\columnwidth]{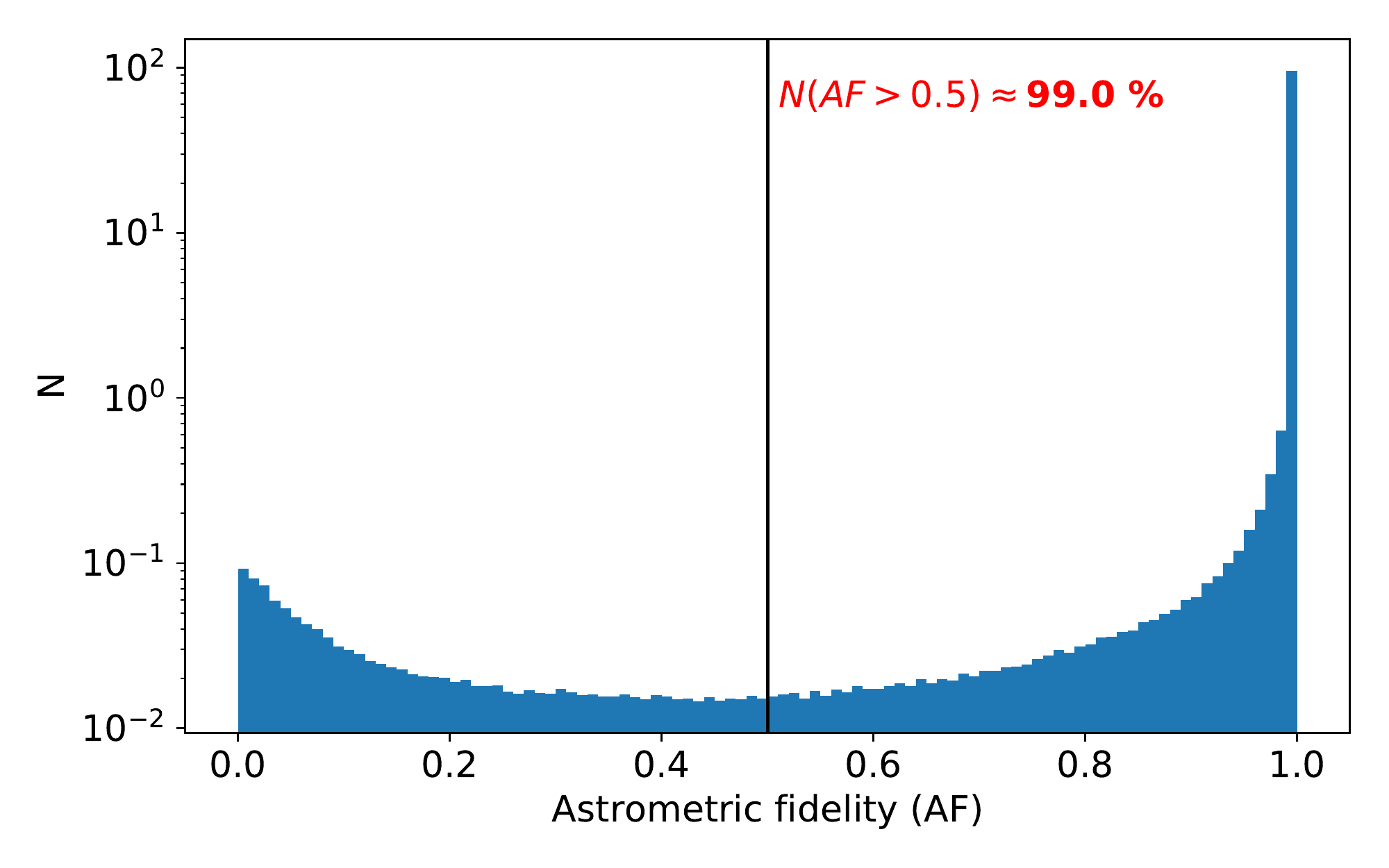} 
\caption{Astrometric fidelity (AF) for the \xrvs{} sample. AF above 0.5 corresponds to good astrometric solutions, and nearly all of our dataset satisfies this condition. \label{fig:fidelity}}
\end{figure}

Despite the high quality of data, systematics and spurious astrometry have been identified in \gaia{} \citep{Fabricius:2021}. These can usually be filtered out using quality parameters such as the re-normalised unit weight error (\ruwe), \astromexnoise{}, \visibperiods{}, and others. \cite{Rybizki:2022} showed that such simple filtering does not necessarily remove spurious sources. They trained a neural network classifier on selected quality filters using high and low signal-to-noise data, and provide 'Astrometric-fidelity (AF)' for the entire \gedrthree{} dataset, such that $ 0$ (bad) $< \rm AF < 1$ (good), with $\rm AF = 0.5$ roughly dividing the two regimes. The distribution of Astrometric-fidelity for our \xrvs{} dataset is shown in \autoref{fig:fidelity}, where over $90\%$ of sources lie above the $\rm AF = 0.5$ threshold and have 'good' astrometry available. For all our analysis we thus discard data with $\rm AF < 0.5$. Furthermore, we restrict our sample to \errorovparallax{} $<0.2$, in order to estimate distances by inverting the parallax ($\varpi$), to which we also apply the recommended zero-point offset of 0.017 mas. Thus after applying our quality filters on the radial velocities, and astrometry, our \xrvs{} dataset consists of $N = 9,407,060$ sources in total. We further restrict this sample to be within a heliocentric distance of 3.5 kpc. We chose this radius as it is a ‘safe limit’ for using inverse parallax distances. This is somewhat arbitrary because one can define the reliable distance error according to their specific science case. In our previous work in K18 we used a similar value so in order to be consistent we chose 3.5 kpc as the maximum radius. More recently, with a sample of red clump stars selected using astrometry from \gedrthree{}, it was shown that the transition beyond which inverting parallax becomes unreliable is at about 4 kpc \citep{gaia_anticenter2021}. So, our distance cut here is safe enough according to such comparisons.. The final number count of stars in our sample is N = 8,448,302.

In order to correctly propagate the uncertainties in the various astrometric quantities, we make use of the set of correlation coefficients between astrometric parameters (ra, dec, parallax, pm\_ra, pm\_dec) provided in \gedrthree{} \citep{Lindegren:2021}. By combining the correlation coefficients (given as `ra\_pmra\_corr', 'dec\_pmra\_corr etc), with the standard uncertainties in each (e.g., 'ra\_error', 'dec\_error' etc), we construct a covariance matrix, and sample 100 times from this using a multivariate normal distribution. The radial velocity uncertainties are also included in the matrix, however, we assume these to have zero correlation with the astrometry.

\subsection{Simulations: phase-mixing of Spiral arms}
\label{sec:toymodel}
The \gaia{} dataset has revealed rich kinematic substructure in the Galaxy. These features have been linked to both internal (due to action of the Galactic bar \& spiral arms) and external (such as accretion events) sources of perturbations. In \citet[][K19 hereafter]{khanna2}, we showed that phase-mixing triggered by disrupting spiral arms can explain some of the the phase space structures discovered with \gdrtwo{} data, in particular, the diagonal ridges visible in the ($R,V_{\phi}$) plane and arches in the ($V_{R},V_{\phi}$) plane. Given that these features are present on large physical scales of several kpc, it is interesting to explore what the signatures of such phenomena are present in the residual velocity field. 
The full details related to the setup of the simulation are given in K19. Briefly, we setup test particles in the configuration of four Archimedean spirals. The radial velocity was sampled from $\mathcal{N}(0, 20)$, and the azimuthal velocity from $\mathcal{N}(\Theta(R),20)$, where $\Theta(R)$ denotes the circular velocity and $\mathcal{N}$ denotes a \textit{Gaussian} distribution. For simplicity, the particles were all setup to be in the plane of the model galaxy, and thus also have zero vertical velocity. A total of 640000 particles were evolved for 650 Myr (in timesteps of 6 Myr) in the \textit{MWPotential2014} potential using the \textit{galpy} package \citep{Bovy:2015}. We compare the simulation to observed data in \autoref{sec:gaia_psd_compare_sim}.

\section{Methods}
\label{sec:methods}
\subsection{Coordinate transformations}
\label{sec:coordtrans}
Throughout this paper, we adopt a right-handed coordinate frame in which the Sun is placed at a Galactocentric distance of $R_{\odot}=8.275$ kpc. This is consistent with the latest ESO Gravity measurement, of the orbit of the star S2 around the Milky Way's supermassive black hole  \citep{GravityCollaboration:2021}. The Sun thus has Galactocentric coordinates $(X,Y,Z) = (-8.275,0,0.25)$ kpc. The cylindrical coordinate angle $\phi={\rm tan}^{-1}(Y/X)$ increases in the anti-clockwise direction, while the rotation of the Galaxy is clockwise. The heliocentric Cartesian frame is related to Galactocentric by $X_{\rm hc}=X+R_{\odot}$, $Y_{\rm hc}=Y$ and $Z_{\rm hc}=Z$. $X_{\rm hc}$ is negative toward $\ell=180^\circ$ and $Y_{\rm hc}$ is positive towards Galactic rotation. For transforming velocities between heliocentric and Galactocentric frames we use $(\dot{X}_{\odot},\dot{Y}_{\odot},\dot{Z}_{\odot})=(U_{\odot},\Omega_{\odot}R_{\odot},W_{\odot})$.
Following \cite{Schonrich:2010}, we adopt $(U,W)_{\odot}=(11.1,7.25)$ \kms{}, while for the azimuthal component we use the constraint of  $\Omega_{\odot}=30.24$ \kms{}kpc$^{-1}$, which is set by the proper motion of Sgr A*, i.e., the Sun's angular velocity around the Galactic center \citep{Reid:2004}. This sets the azimuthal velocity of the Sun to $V_{\phi,\odot}=-250$ \kms{}, rounded to three significant figures. 
\begin{figure*}
\includegraphics[width=2.\columnwidth]{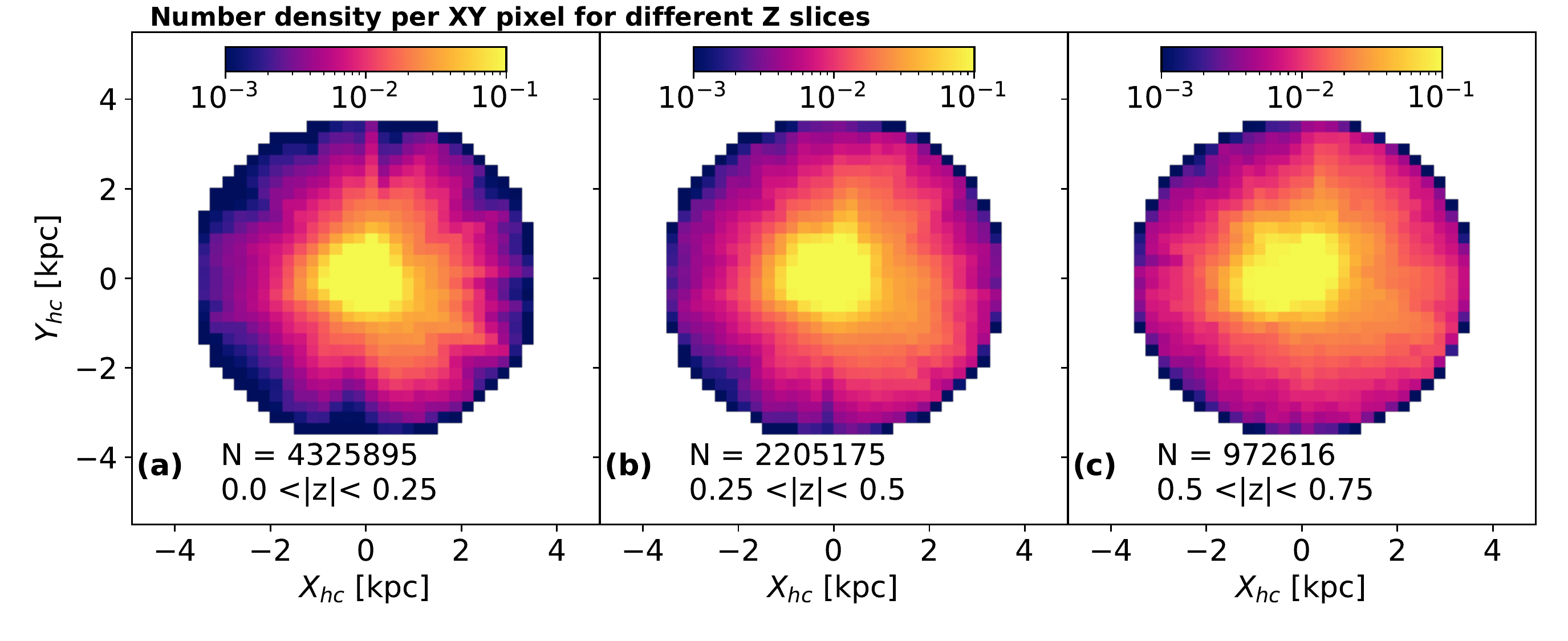} 
\caption{Stellar number density \textbf{(normalised)} shown for the three main slices in \mnz{} used in our analysis. All data here is within 3.5 kpc from the Sun. The Galactic Center is placed at $X_{\rm hc},Y_{\rm hc}$=($8.275,0$).\label{fig:density_slices}}
\end{figure*}

\subsection{Axisymmetric model and \textbf{residual} velocity maps}
\label{sec:psd_gaia_kinmod}

Our goal is to quantify the deviations in the motion of individual stars, with respect to a mean ordered 3D velocity field. To achieve this, we first construct a simple axisymmetric model, assuming that the mean rotational motion, \vphimod{}, is a function only of \rad{}, and \mnz{}. Additionally, we also assume equilibrium in the other two components, such that, \vRmod{} = 0, and \vzmod{} = 0. The \vphimod{}, can be constructed in two different ways, a) by dividing the data into thin vertical slices in $z$, and interpolating the rotation curve as a function of \rad; or alternatively, b) we can follow the approach as in \khannalos{}, where we take into account the vertical gradient in the azimuthal velocity, by fitting a 2D polynomial, such as, 

%
\begin{equation}
\label{eqn:vphi_model}
V_{\rm \phi, mod} (R,Z)  = {\sum\limits_{i=0}^{2}}{\sum\limits_{j=0}^{2}}  a_{ij}  (R-R_{\odot})^{i}Z^{j}.
\end{equation}

We follow this approach of fitting a global model to our dataset, using the \textit{scipy.optimize.curve\_fit}\footnote{\url{https://docs.scipy.org/doc/scipy/reference/generated/scipy.optimize.curve_fit.html}} module. To estimate the uncertainties, we perform the fit over multiple realisations of the \xrvs{} dataset. This Galactocentric model is then transformed to the heliocentric
frame, in order to obtain a model line-of-sight velocity component, again using $(\dot{X}_{\odot},\dot{Y}_{\odot},\dot{Z}_{\odot})=(U_{\odot},\Omega_{\odot}R_{\odot},W_{\odot})$, as described in \autoref{sec:coordtrans}.

\begin{figure*}
\includegraphics[width=1.7\columnwidth]{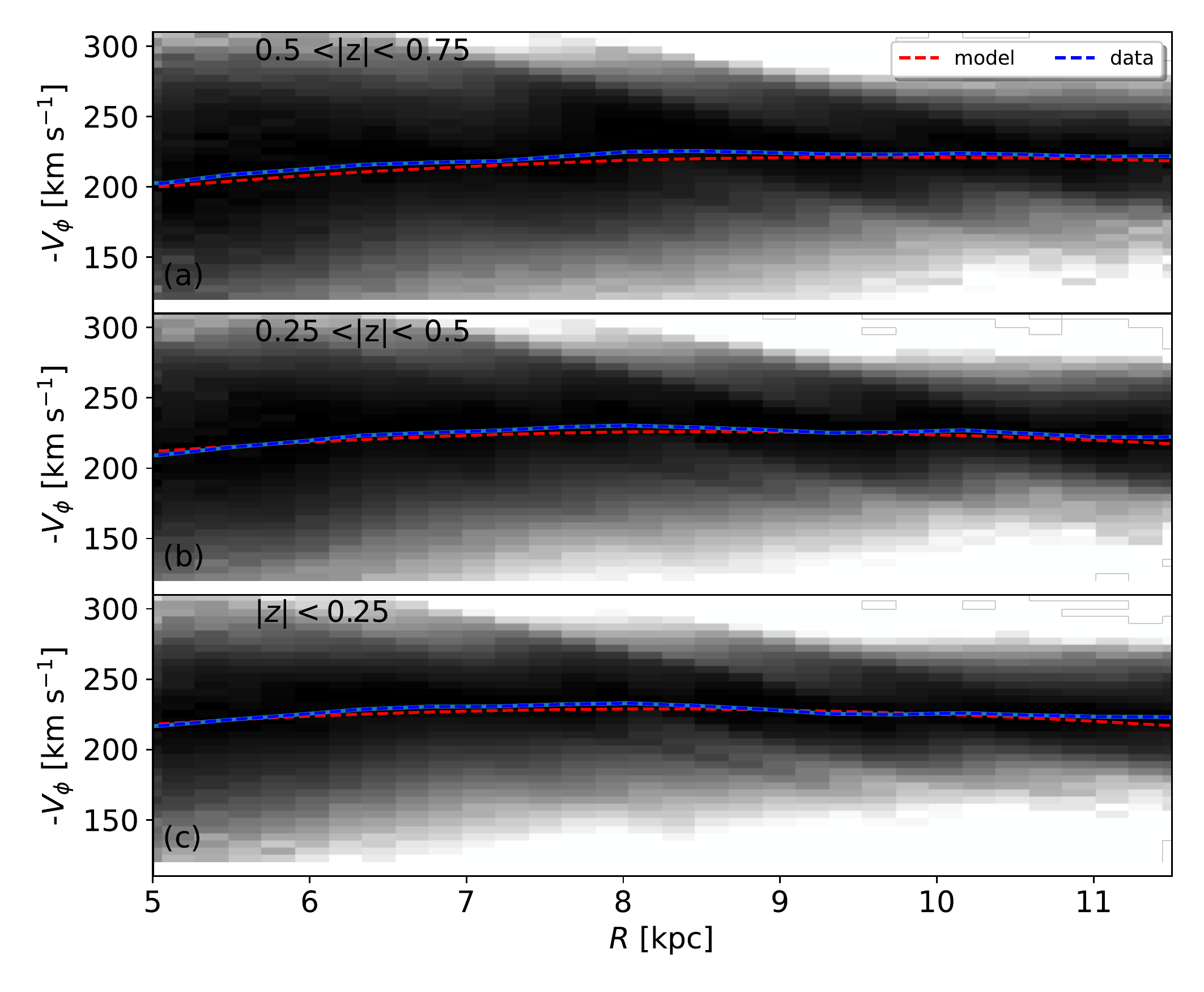} 
\caption{The global $R$-\vphi{} distribution for our dataset, shown for each \mnz{} slice separately. The mean of the distribution (shown in blue) is fitted with a polynomial (shown in red) that takes into account the radial and vertical profile of the rotation curve. \label{fig:rotcurve}}
\end{figure*}

Thus, we can subtract off for each star, $i$, the mean motion in the four velocity components, $V_{\phi,R,z,\rm los}$, in order to obtain the residual velocity, $\Delta V_{i} = V_{i} - V_{i, {\rm mod}}$. We divide the data into three slices in \mnz{}, namely, a) \midslice{}:
$|z|/[{\rm kpc}] < 0.25$, b) \upperslice{}: $0.25 < |z|/[{\rm kpc}] < 0.5$, and, c) \topslice{}: $0.5 < |z|/[{\rm kpc}] < 0.75$. In each \mnz{} slice, then, we present maps of the residual velocity field, by binning up stars in the $(x,y)$ plane, with a resolution of $0.25\times0.25$ kpc$^{2}$. In order to minimise the Poisson noise, we require a minimum number of 100 stars per pixel, and call this quantity, $N_{minbin}$. This ensures that the standard error in the mean velocity is well below 1 \kms{} per pixel. Our data slicing, binning, and coordinate system, are illustrated in the number density maps shown in \autoref{fig:density_slices}.

\subsection{Fourier analysis}
\label{sec:fourier_method}

In order to further characterise the non-axisymmetric motion in the Galaxy, we perform Fourier analysis on the residual velocity maps. If $A_{\rm kl}$ is the 2D Fast Fourier Transform (FFT) of the image $h$ (our maps in xy plane) and $\Delta x$ and $\Delta y$ are the size of the bins along the $x$ and $y$ directions, the 2D power spectrum of the residual velocity field is then given by 
\begin{equation}
P_{kl} = \frac{1}{N_{\rm eff}} |A_{\rm kl}|^{2} \Delta x \Delta y .      
\end{equation}
Here $N_{\rm eff}=\sum_{i} \sum_{j} H (n_{ij}-100)$ is the effective number of bins in the image, where $H$ is the \textit{Heaviside} step function and $n_{ij}$ is the number of stars in the $(i,j)$-th bin. Next, we average $P_{kl}$ azimuthally in bins of $k=\sqrt{k_x^2+k_y^2}$, to obtain the 1D power spectrum $P(k)$. The $P(k)$, as defined above satisfies the following normalization condition given by the \textit{Parseval's} theorem,
\begin{eqnarray}
\int_0^{\infty} \! P(k) 2 \pi k  \, \mathrm{d}k &=& \sum_k \sum_l P_{kl} \Delta k_x \Delta k_y \\ \nonumber
&=& \frac{\sum_{i} \sum_{j} H (n_{ij}-100) h_{ij}^2}{N_{\rm eff}} .
\end{eqnarray}

Here, $h_{ij}$ is the residual velocity in the ($i,j$)-th bin. $\sqrt{P(k)}$ has dimensions of \kms{} and denotes the amplitude of the fluctuations.
The presented formalism ensures that the estimated power spectrum $P(k)$ is invariant to the changes in the sizes of the bins, to the changes to the overall size of the image box, and to bins with missing data. Finally, we locate the coordinates ($k,\sqrt{P(k)} $) of the peak(s) in our power spectrum, using the \textit{scipy.signal.find\_peaks}\footnote{\url{https://docs.scipy.org/doc/scipy/reference/generated/scipy.signal.find\_peaks.html}} routine.

\textbf{Noise}: To illustrate the robustness of our results, we also compute the power spectrum of a residual velocity field consisting purely of noise. For this, we generate 100 random images, where, the value in each pixel of an image is sampled from a normal distribution with zero mean and dispersion equal to one standard deviation of the residual velocity ($\Delta V_{i}$) in each pixel of the real image. Following this, we compute the power spectrum for each noise image, and from the sample, we obtain the power at the 50$^{th}$ (median), $84^{th}$, and $16^{th}$ percentiles, and thus the spread due to noise.

\textbf{Uncertainties}: Propagating uncertainties from the physical to the Fourier space is not a trivial exercise. So, in order to estimate the uncertainty in the amplitude $\sqrt{P(k)}$, we compute the power spectrum for 20 random data realisations, sampled from a multivariate \textit{Gaussian} distribution with covariance matrix given by the errors and correlation from the \gedrthree{} catalogue, as described in 
\autoref{sec:datasets}. From this, we again use the power at the 50$^{th}$ (median), $84^{th}$, and $16^{th}$ percentiles, and thus the spread due to uncertainties in our observables.

\section{Results}
\label{sec:gaia_psd_results}
\subsection{\xrvs{} dataset }

\begin{figure*}
\includegraphics[width=2.\columnwidth]{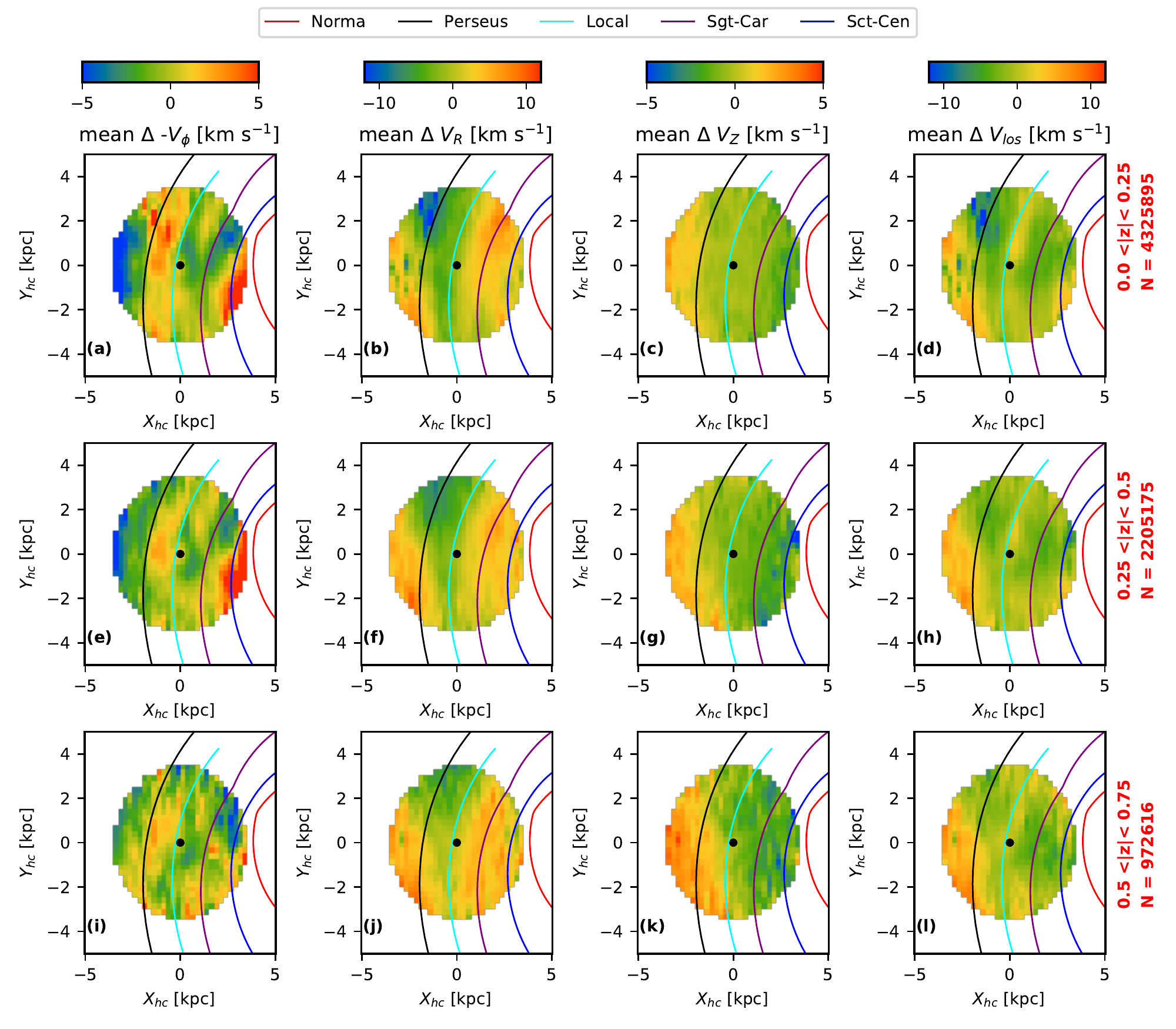} 
\caption{Heliocentric map of velocity residuals, $\Delta V_{i}$, in the four components, $V(\phi,R,z,\rm los)$ for the combined ($|z|$) slices. Panels (a-d) show the maps for the \midslice{}, i.e., $|z|/[{\rm kpc}] < 0.25$, b) Panels (e-h) show the \upperslice{}, i.e., $0.25 < |z|/[{\rm kpc}] < 0.5$, and, c) Panels (i-l) show the \topslice{}, i.e., $0.5 < |z|/[{\rm kpc}] < 0.75$. For each $|z|$ slice, the number of stars used is also indicated. The solid lines overplotted on the maps represent model of Spiral arms taken from \citet{Reid:2019}. The Galactic Center is placed at $X_{\rm hc},Y_{\rm hc}$=($8.275,0$). \label{fig:cmb_vmap}}
\end{figure*}

\subsubsection{Residual velocity maps}

We first apply our method to slices in \mnz{}, i.e., where data in symmetric slices (above and below the Galactic plane) has been combined. The best fitting parameters for \autoref{eqn:vphi_model} are listed in \autoref{app:best_fit}, and the fit itself is illustrated in \autoref{fig:rotcurve}, for the individual \mnz{} slices. In \autoref{fig:cmb_vmap}, we show the 2D heliocentric ($xy$) maps of velocity residuals ($\Delta V_{i} $) in the four components, i.e., \vall{}. \autoref{fig:cmb_vmap}(a-d) shows the maps for the \midslice{}, \autoref{fig:cmb_vmap}(e-h) shows the \upperslice{}, and, \autoref{fig:cmb_vmap}(i-l) shows the \topslice{}. Moving from top to bottom in this scheme is thus akin to moving away from the Galactic plane.

First, looking at the the \midslice{}, it is immediately clear that the residual velocity field is not smooth across at least three components, i.e., $V_{\phi,\rm R,\rm los}$. There are regions of high $|\Delta V_{i} |$, that span across several pixels, and translate to a physical scale of 100s of parsecs. The magnitude of the $|\Delta V_{i} |$ is of the order of $10$ \kms{}, in the $V_{\rm R,\rm los}$ components, and about $5$ \kms{} in the $V_{\phi}$ component. On the other hand, \autoref{fig:cmb_vmap}(c) shows that the \vz{} component, shows a smooth gradient, reaching about $\Delta V_{\rm z} = 5$  \kms{}, in the outer disc. In addition to the large scale patterns seen in the maps, we also note the presence of a few prominent pixels with very high $|\Delta V_{i} |$, for example, such as near ($X_{\rm hc}=-0.9,Y_{\rm hc}=1.2$), and ($X_{\rm hc}=-2.5,Y_{\rm hc}=0$), in \autoref{fig:cmb_vmap}(a,b,d). In order to check if these pixels were not merely a result of \textit{ Poisson} noise, we present a similar map for the \midslice{} again in \autoref{app:scattered_pixels}, however, with $N_{minbin} = 500$. The peculiar pixels are clearly still present even after reduced \textit{Poisson} noise.

For the intermediate region, i.e., \upperslice{}, the features are largely unchanged compared to the \midslice{}. The $\Delta V_{\phi} $ component in \autoref{fig:cmb_vmap}(e) appears smoother compared to \autoref{fig:cmb_vmap}(a). The large scale features in the $\Delta V_{R} $ component, in \autoref{fig:cmb_vmap}(f) are also not too dissimilar to that in \autoref{fig:cmb_vmap}(b), however, we do note the absence of the peculiar pixels seen in the \midslice{}. In the $\Delta V_{z}$ component, shown in \autoref{fig:cmb_vmap}(g), the positive residuals towards the outer disc are more prominent compared to \autoref{fig:cmb_vmap}(c). Overall, the maps in the \upperslice{} retain the large scale features seen in the \midslice{}, but appear smoother.

Finally, for the \topslice{}, in \autoref{fig:cmb_vmap}(i-l) we notice that the peculiar pixels are completely absent, and the velocity residuals appear spread over an even larger area compared to the two lower $|z|$ slices. In all three $|z|$ slices, the $\Delta V_{\rm los} $ map shows a superposition of features in the other three components $V_{\phi,\rm R,\rm z}$. This is not surprising, given that the other components all contribute towards the line-of-sight velocity. As a consequence of this, the component with the highest velocity residuals dominates the $\Delta V_{\rm los} $ map.

Using parallaxes and proper motions of masers, from the \textit{BESSEL} survey, \cite{Reid:2019} built a \textit{log-periodic} model of the Galactic spiral arms. Using the model (their Table 2.), in each panel in \autoref{fig:cmb_vmap}, we overplot the heliocentric positions of the main arms: Norma-Outer (\textit{Norma }hereafter), \textit{Perseus}, \textit{Local}, Sagittarius-Carina arm (\textit{Sgt-Car} hereafter), and Scutum-Centaurus-OSC arm (\textit{Sct-Cen} hereafter). We ignore the thickness of the arms, as this would make the maps difficult to read. The mean locations of the arms shows that the \textit{Norma} arm (in red) lies just outside the extent of our data selection. There is however, a significant overlap between the other four spiral arms and the data. In particular, we note in \autoref{fig:cmb_vmap}(b,f,j), the presence of a strong gradient in the $\Delta V_{R} $ component about $Y_{\rm hc}=0$, that lies exactly at the location of the \textit{Perseus} arm (in black). The gradient gets weaker with $|z|$, but is still present in the highest slice. Interestingly, the positive residuals in $\Delta V_{z}$, also line up well with the \textit{Perseus} arm, in \autoref{fig:cmb_vmap}(c,g,k). In the region covered by the \textit{local} arm, the $\Delta V_{R} $ component is on average negative, i.e., stars seem to be moving inwards w.r.t the Galactic Center. Meanwhile, in the region covered by the \textit{Sgt-Car} arm, the the $\Delta V_{R} $ component is on average positive. Finally, the resultant $\Delta V_{\rm los} $ component also has a positive feature near the location of the \textit{local} arm.

\subsubsection{Power spectrum}

\begin{figure*}
\includegraphics[width=2.\columnwidth]{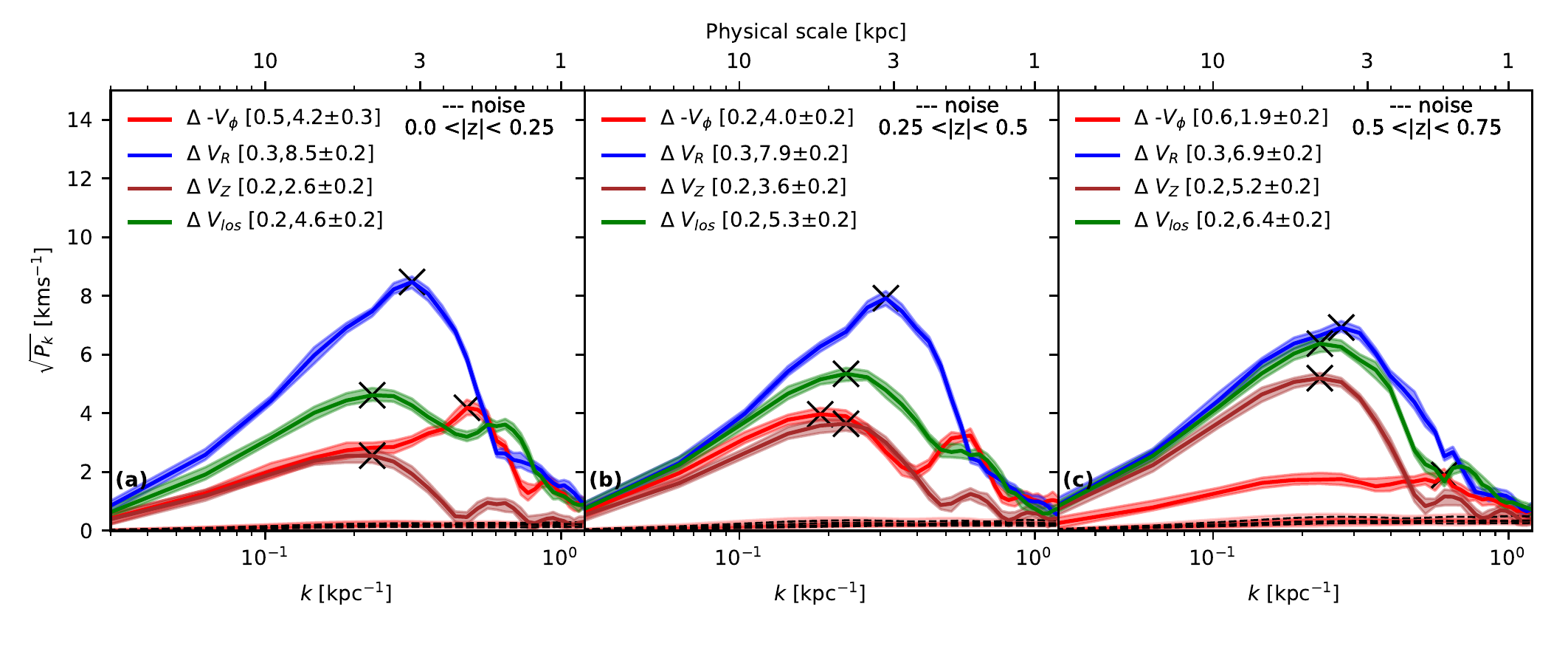} 
\caption{Power spectrum of velocity residuals, $\Delta V_{i}$, for the combined ($|z|$) slices, corresponding to the kinematic maps shown in \autoref{fig:cmb_vmap}. The quantities in brackets denote the coordinates (wavenumber $k$, power $\sqrt{P(k)}$) of the peak in each component, marked with a black cross.  In the \midslice{}, the highest power is in the \vR{} (radial) component, and the lowest is in the \vz{} component. As we go above the plane, the amplitude drops in all components except for in the \vz{} component. The radial streaming motion dominates in all $|z|$ slices. The profile with the dashed line shows the power spectrum expected due to noise, and is essentially negligible. \label{fig:cmb_psd}}
\end{figure*}

Next, in \autoref{fig:cmb_psd}, we show the power spectrum of the $\Delta V_{i}$ in each velocity component, and for each of the three $|z|$ slices. For each velocity component, we mark the position of the `peak' in the power spectrum, with a black cross. Additionally, the coordinates ($k, \sqrt{P(k)}$) of the peaks, along with the $1\sigma$ uncertainty is also included in the plot labels. The typical $1\sigma$ uncertainty on $\sqrt{P(k)}$ is of the order of 0.2 \kms{}. Lastly, the top axes in each panel plots the quantity, $1/k$, in order to give a sense of the physical scale associated with the peaks in the power spectrum.

Looking at the the \midslice{} first, and considering only the values of the peaks in the power spectrum, \autoref{fig:cmb_psd}(a) shows that the peak is highest in the \vR{} component ($\sqrt{P(k)} = 8.5 $ \kms), and lowest in the \vz{} component ($\sqrt{P(k)} = 2.6 $ \kms). The peak in the \vlos{} component 
is smaller than the dominant component (\vR{}), with a value of $\sqrt{P(k)} = 4.5 $ \kms. Lastly, the peak in the \vphi{} component, is at $\sqrt{P(k)} = 4.2 $ \kms. With the power spectrum we also have the location of the peaks in $k$ or real ($1/k$) space. In the \midslice{}, most of the power is concentrated between wavenumbers, $ 0.2 < k < 0.4 $, corresponding to physical scales between, $5 > 1/k > 2.5$ kpc.

In the \upperslice{}, in \autoref{fig:cmb_psd}(b), the patterns are not too dissimilar from the \midslice{}, in that, once again, the peak in the power spectrum is the highest in the \vR{} component ($\sqrt{P(k)} = 7.9 $ \kms). The power in the \vphi{} component ($\sqrt{P(k)} = 4.0 $ \kms), is very slightly lower than that in the \midslice{}, but the location of the peak has shifted slightly towards lower wavenumbers. However, given the broad distribution of these peaks, we caution against using the peak locations as strict values. In general, though, the power in the \vR{} component is still dominant and more concentrated in this intermediate $|z|$ slice. It is interesting to note, also, that the values of the peaks have dropped for the \vR{} and \vphi{} components, while increasing in the \vz{} component. Furthermore, the peak in the \vlos{} has also increased by 0.7 \kms{}. This is most likely a consequence of the combination of the high power in both \vR{} and \vz{} contributing towards the \vlos{} component.

\begin{figure}
\includegraphics[width=1.\columnwidth]{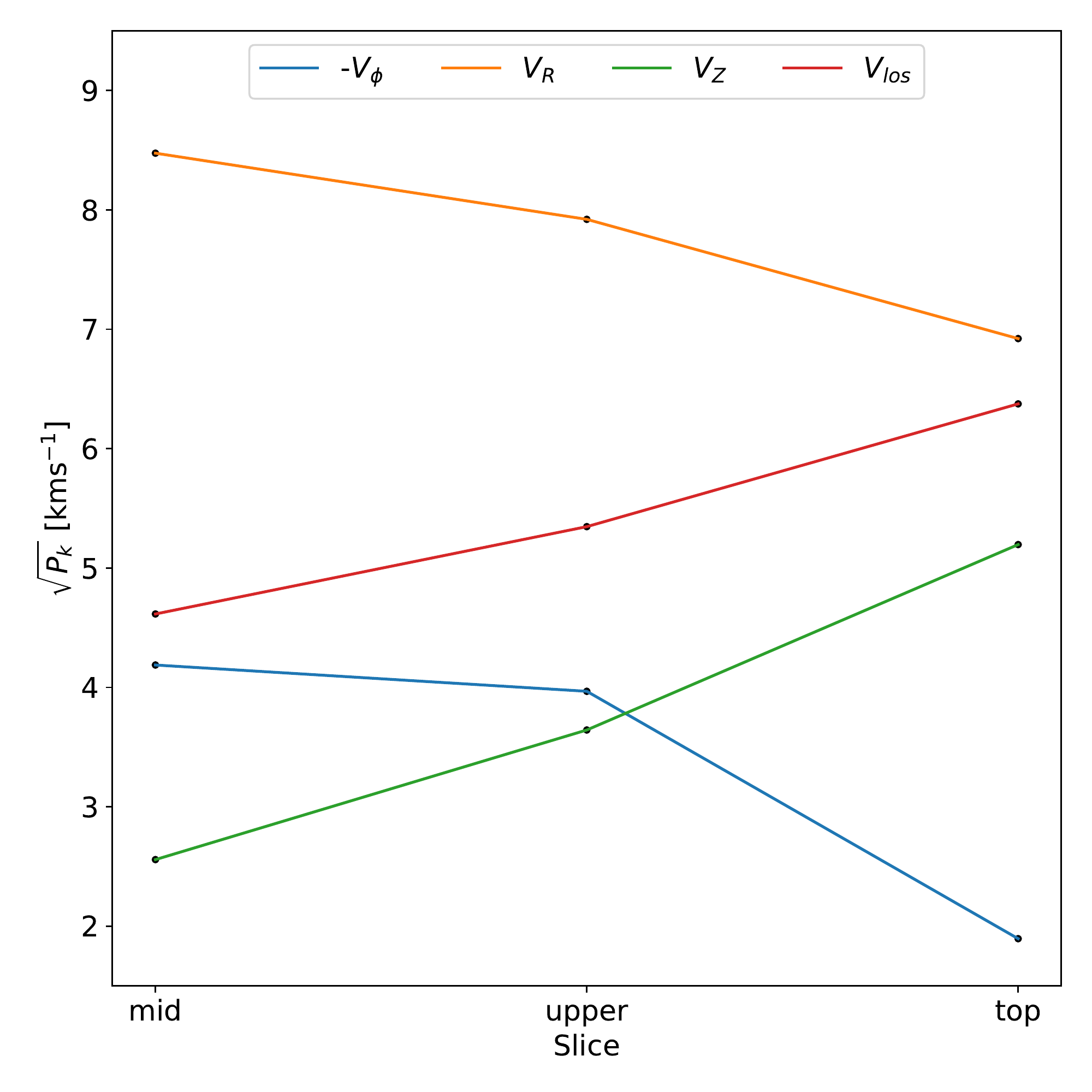} 
\caption{Peak power of velocity residuals ($\Delta V_{i}$) as a function of height above the Galactic plane, represented here by \mnz{} slices. In general, the power in the radial and the azimuthal components drops with an increase in height above the Galactic plane, while that in the vertical component increases. \label{fig:cmb_psd_z}}
\end{figure}

Finally, for the \topslice{}, \autoref{fig:cmb_psd}(c) shows that while the peak power is still the highest in the \vR{} component ($\sqrt{P(k)} = 6.9 $ \kms), the peak in the \vz{} component is now at its maximum ($\sqrt{P(k)} = 5.2 $ \kms). Furthermore, the distribution of the power spectrum in the \vz{} component is also more concentrated compared to the lower two \mnz{} slices. The power in the \vphi{} component has reduced to $\sqrt{P(k)} = 2.1 $ \kms, and the profile is flat across $k$. The \vlos{} component is highest ($\sqrt{P(k)} = 6.4 $ \kms) compared to the lower two slices, and follows closely the amplitude of the \vR{} and the \vz{} components. The evolution of the peak power in each component with height above the plane is illustrated in \autoref{fig:cmb_psd_z}. 
Additionally, we summarise the characteristics of the power spectra in \autoref{tab:psd_stats}, in terms of the width of the peaks. For each slice and velocity component, we compute a full-width-at-half-maximum (FWHM) like measure, except at 70\% of the peak instead of 50\%, as the peaks are not Gaussian like.  In a few of the components there are two significant peaks, so for these, both peaks are considered while providing the range.

\subsection{Comparison with phase-mixing simulation}
\label{sec:gaia_psd_compare_sim}

\begin{figure*}
\includegraphics[width=2.\columnwidth]{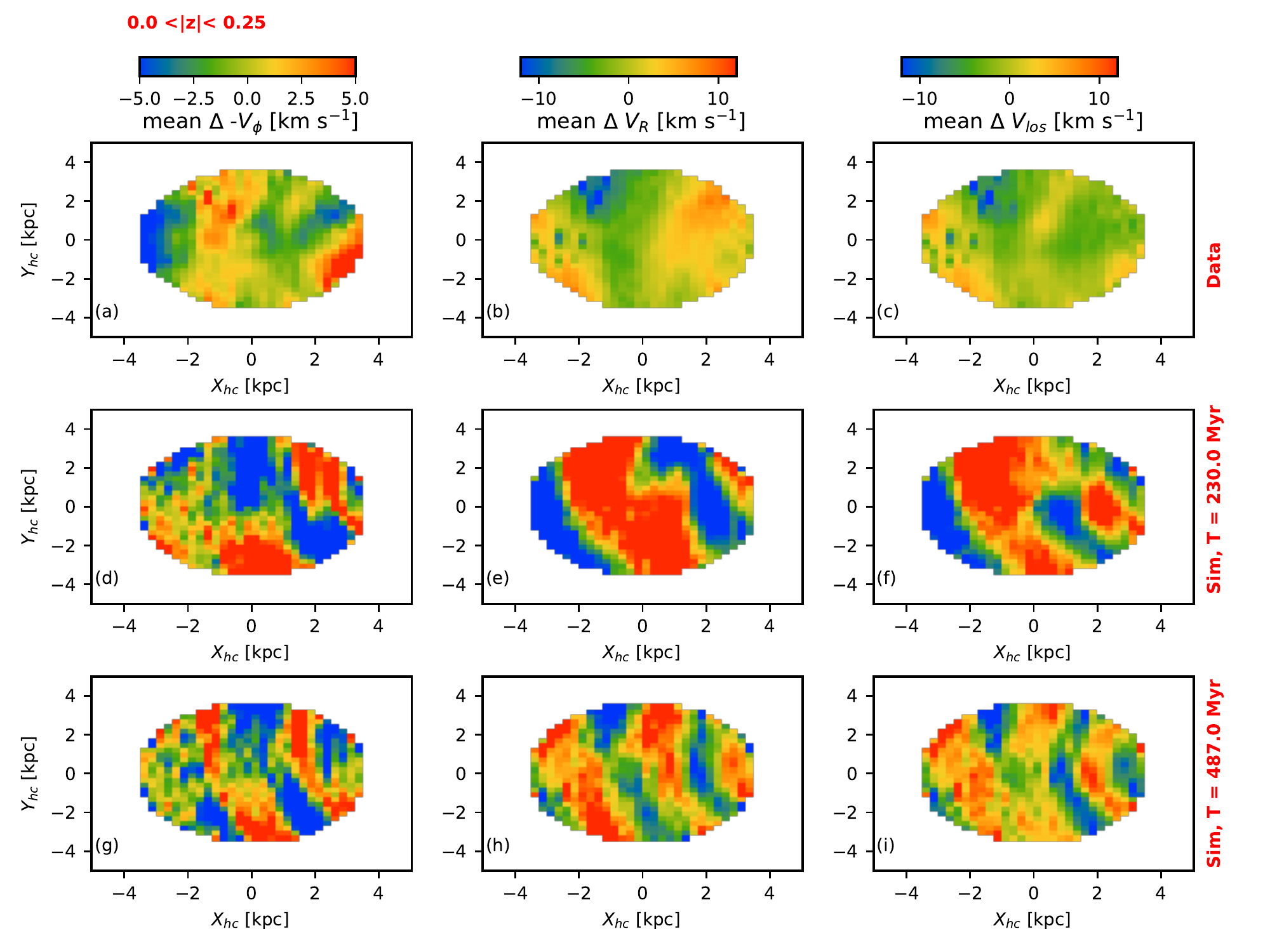} 
\caption{Heliocentric map of velocity residuals, $\Delta V_{i}$, in the components, $V(\phi,R,\rm los)$, compared between the observed data in the $|z|/[{\rm kpc}] < 0.25$ slice (panels a-c), and our Phase-mixing simulation from \citet{khanna2}, at $\tau = 230$ Myr (panels d-f), and at $\tau = 487$ Myr (panels g-i). In all three cases, the residuals are highest in the $\Delta V_{R}$ component. At later stages in the simulation, the $\Delta V_{i}$ pattern seems to appear like that in the observed data. \label{fig:cmb_vmap_sim_comparison}}
\end{figure*}

We now use the simulation run by K19 to explore the residual velocity field in the scenario where disrupting spiral arms is a source of perturbation, and phase-mixing as the Galaxy relaxes, produces the kinematic structures such as ridges and arches seen in \gdrtwo{}. We carry out the analysis described in \autoref{sec:methods}, in exactly the same manner as for the observed data, except that since our simulated particles are all confined to the plane, we limit our comparison to the \midslice{} only, and likewise, since there is no vertical motion in the simulation, we only consider the three components $V(\phi,R,\rm los)$. For each timestep, we produce maps of velocity residuals similar to \autoref{fig:cmb_vmap}. Our simulation is by no means supposed to be a one-to-one comparison with the Milky Way, but for the purpose of a simple comparison, we handpick two snapshots, one at an early stage (230 Myr), and another at a later stage (487 Myr) of the phase-mixing process. In \autoref{fig:cmb_vmap_sim_comparison}, we present the $\Delta V_{i}$ maps for these two snapshots, and compare it to the \midslice{} maps in the data. \autoref{fig:cmb_vmap_sim_comparison}(a-c) are a replica of \autoref{fig:cmb_vmap}(a,b,d), while \autoref{fig:cmb_vmap_sim_comparison}(d-f) shows the $\Delta V_{i}$ at 230 Myr. At this early time, there are regions with very high $-\Delta V_{i}$ and $+\Delta V_{i}$ residuals present in the three components, and these manifest as broad diagonal features. Interestingly, the radial ($\Delta V_{R}$) component seems to dominate in the simulation as well, and as in the data, the $\Delta V_{los}$ component follows most of the pattern of the radial component. In comparison, at later times (487 Myr), \autoref{fig:cmb_vmap_sim_comparison}(g-i) shows that the residuals are now confined in much thinner stripes, and the amplitude also seems lower. Notwithstanding the caveat that our simulation is not selection-function matched, our crude qualitative comparison, does however show that the diagonal residuals seen in the data have a similar appearance to that noted in \autoref{fig:cmb_vmap_sim_comparison}(g-i).

\begin{table}
\centering
\caption{Power spectra summary. For each velocity component, we tabulate  the range in wavenumber ($k$), over which 70\% of the power is concentrated. In cases where there are multiple peaks present, the range including the second peak is also provided. \label{tab:psd_stats}}
\begin{tabular}{l|l|l|l|l}
\hline
  & \vphi{} & \vR{} & \vz{} & \vlos{}  \\
\midslice{}  & [0.27, 0.62] & [0.11, 0.71] & [0.15, 0.49] & [0.11, 0.31]    \\
& & & [0.15, 0.74] &  [0.11, 0.86] \\
\upperslice{}  & [0.10, 0.64] & [0.11, 0.41] & [0.16, 0.49] & [0.10, 0.63]    \\
& & & [0.16, 0.74] & \\
\topslice{}  & [0.11, 0.74] & [0.11, 0.42] & [0.13, 0.45] & [0.11, 0.35]    \\
& & & & \\
\hline
\end{tabular}
\end{table}

\begin{figure*}
\includegraphics[width=2.\columnwidth]{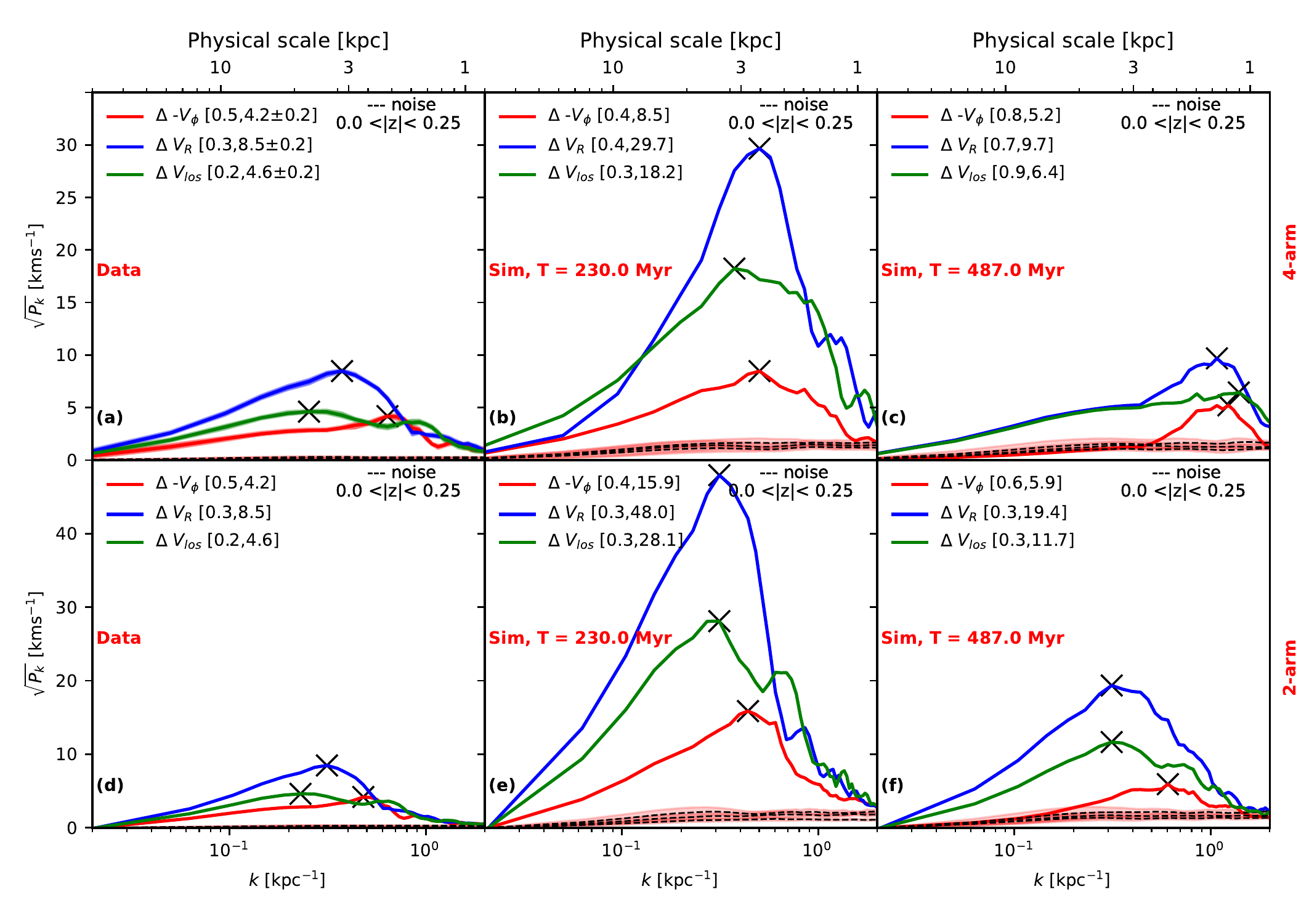} 
\caption{Power spectrum of velocity residuals, $\Delta V_{i}$, compared between the observed data in the $|z|/[{\rm kpc}] < 0.25$ slice (panels a,d), and our Phase-mixing simulation from \citet{khanna2}, at $\tau = 230$ Myr (panel b), and at $\tau = 487$ Myr (panel c). The quantities in brackets denote the coordinates (wavenumber $k$, power $\sqrt{P(k)}$) of the peak in each component, marked with a black cross. At early times in the simulation, the peak in the power spectrum have a very high amplitude, which drops at later times. Moreover, the peaks shift towards smaller physical scales with time. Panels (e-f), are similar to panels (b-c), except this is shown for a 2-arm spiral model. In such a 2-arm configuration, the shift in the peak power is smaller compared to that in the 4-arm model. In all cases, however, the the highest power is in the \vR{} (radial) component.  \label{fig:cmb_psd_sim_comparison}}
\end{figure*}

As was done for the observed data, we also produce a power spectrum of $\Delta V_{i}$ for the selected snapshots. \autoref{fig:cmb_psd_sim_comparison}(a,d) reproduce the power spectrum of the data in the \midslice{}, \autoref{fig:cmb_psd_sim_comparison}(b,e) presents the power at 230 Myr, and \autoref{fig:cmb_psd_sim_comparison}(c,f) at 487 Myr in the simulation. At 230 Myr, \autoref{fig:cmb_psd_sim_comparison}(b) shows that the power is dominated by the $\Delta V_{R}$ component (peak at $\sqrt{P(k)} = 29.7 $ \kms), and is much higher than the peak power in the $\Delta V_{\phi}$ component ($\sqrt{P(k)} = 8.5 $ \kms). At 487 Myr, \autoref{fig:cmb_psd_sim_comparison}(c), the peaks have dropped to $\sqrt{P(k)} = 9.7 $ \kms{} in the $\Delta V_{R}$ component, and to $\sqrt{P(k)} = 5.2 $ \kms{} in the $\Delta V_{\phi}$ component. Furthermore, the peaks seem to have shifted to smaller physical scales, for example in the $\Delta V_{R}$ component, the peaks shift from $1/k=2.5$ kpc (at 230 Myr), to $1/k= 1.4$ kpc (at 487 Myr). Our simulation is setup as a 4-arm spiral Galaxy, however, we can also restrict to using only a 2-arm setup. The results in this scenario are presented in \autoref{fig:cmb_psd_sim_comparison}(e-f). This seems to have two clear effects, i.e., at a given snapshot a) the amplitudes of the peak power in each component is higher in the 2-arm model, b) the peaks shift slightly onto larger physical scales. The physical scale of the peaks at 487 Myr, in the 2-arm model, are a closer match to the observed data.

\begin{figure*}
\includegraphics[width=2.\columnwidth]{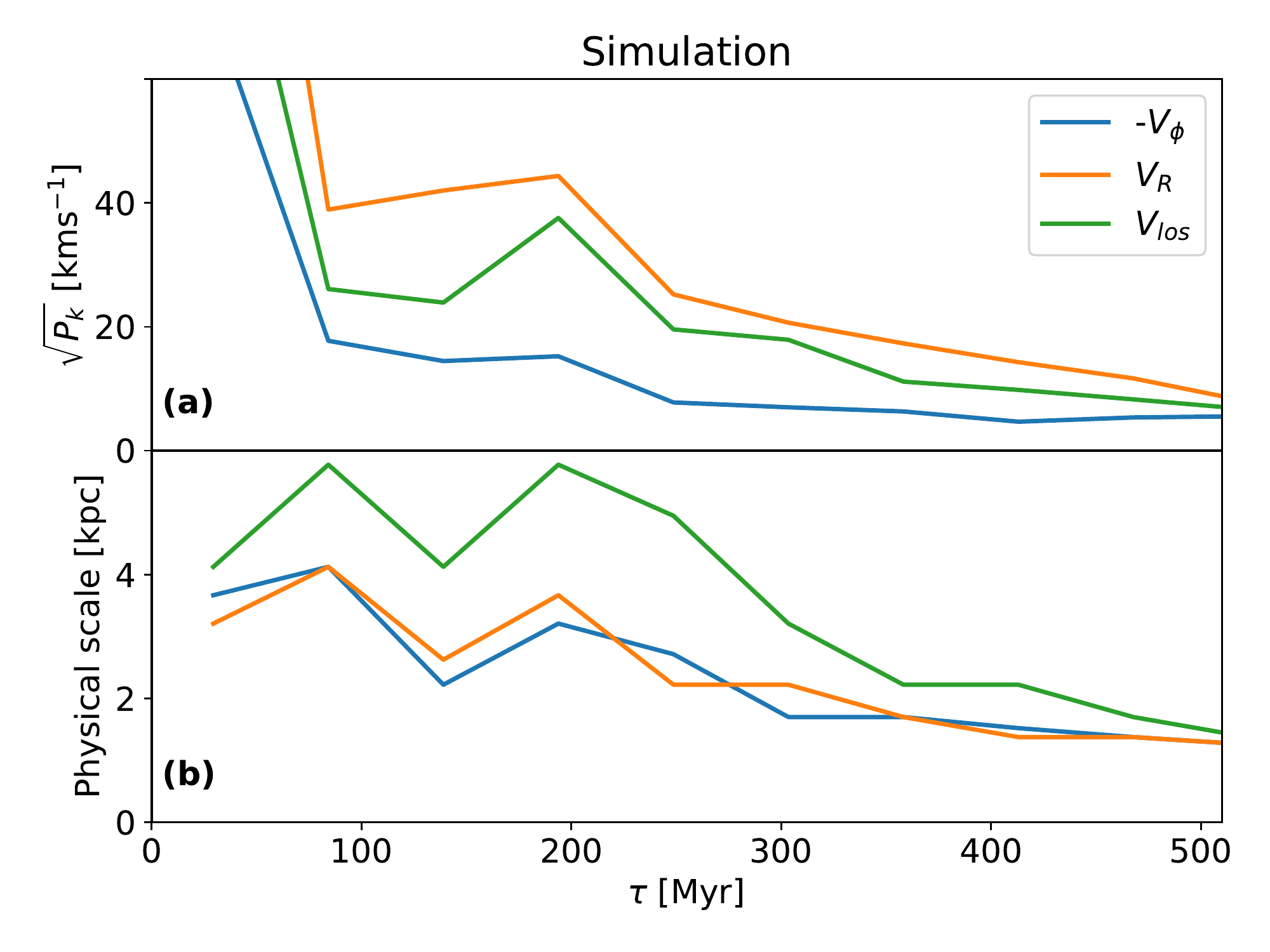} 
\caption{Time evolution of peak power (panel a), and the corresponding physical scale (panel b) in the components $V(\phi,R,\rm los)$ for the phase-mixing simulation. Panel (a) shows that the peak in each component drops with time, and that the \vR{} component dominates at all times. Panel (b) shows that the physical scale of the peaks shifts from higher to lower values with time. A time interval of about 50 Myr is used in the plots. Essentially, as the mock Galaxy relaxes, the streaming motion occurs on smaller scales and with smaller amplitude.\label{fig:time_eval}}
\end{figure*}

Lastly, in \autoref{fig:time_eval}, we plot the peak power (panel a) and the corresponding physical scale (panel b) in the components $V(\phi,R,\rm los)$, at each snapshot through the simulation. Overall, as the mock Galaxy relaxes following the initial perturbation, the peak in the power spectrum drops to lower amplitudes, and also shifts towards smaller physical scales with time. Detailed simulations of the Galaxy could perhaps match such properties of the current measured power spectra in order to time various perturbation events in the Milky Way's history.

\section{Discussion}

We have explored the non-axisymmetric motions of stars out to 3.5 kpc from the Sun, using a dataset with high-precision 6D phase-space. The maps of velocity residuals reveal large bulk flow in the Galactocentric Radial direction, and also non-negligible bulk motion in the azimuthal and the vertical directions. This is the first time the velocity residuals in a large volume have been quantified in the individual components. This is thanks to the availability of high precision proper motion and line-of-sight velocities from the \gaia{} survey. In a similar vein to ours, recently  \cite{Martinez-Medina:2022} also studied the velocity residuals with respect to an axisymmetric model for the rotation curve (albeit a much simpler linear model). Their kinematic maps showed clear correlation between high residuals in the  $\Delta V_{\phi}$ component, and trace of the Milky Way's spiral arms. In the  Pre \gaia{}, 6D velocity mapping of the disc was limited to smaller samples, due to low proper motion precision. Despite this, such studies already hinted at large scale streaming motion in the disc \citep[][etc]{Widrow:2012,Carlin:2013}. Mapping the large scale kinematics of the disc was mostly restricted to the line-of-sight velocities to deduce the global motion. In K18, using Red Clump giants, we showed that the peak power in the $\Delta V_{\rm los}$ component was about 6.3 \kms{} in the \midslice{}. This value seems to be a little higher than our new estimate of 4.6 \kms{}, as shown in \autoref{fig:cmb_psd}(a). However, we can now also note that the line-of-sight velocities do not tell the complete story, and that the residuals are very high in the Galactocentric radial direction. Away from the plane, in K18 we found the peak power in $\Delta V_{\rm los}$ to be about 5 \kms{}, however, due to the sparse data coverage, our measurements were largely consistent with noise. \autoref{fig:cmb_psd} shows that the residuals in the line-of-sight velocity in the \midslice{} and the upper slices only differ by about 1 \kms{}, but thanks to the resolved components, we can now understand why this is the case. The contribution from the azimuthal streaming motion decreases with an increase in height above the plane, but this is compensated by the increase in residuals in the vertical component, while the radial component does not decrease sharply. Since \tgas, there has been evidence for such vertical waves, where the \vz{} component was seen rising as a function of radius and angular momentum $L_{z}$ \citep{Schonrich:2018,huang:2018}. The monotonic rise in \vz{} is considered to mark the onset of the stellar warp, and subsequent data releases have studied in detail, the substructure associated with this trend \citep{Poggio:2018,antojaAC,McMillan:2022,Drimmel:2022}. Furthermore, since the outer disc is flared, the $|z|$ for the stars here is also higher. This would explain the increasing \vz{} residuals with Galactic height in \autoref{fig:cmb_vmap}(c,g,k). 

\subsection{The Local Standard of rest}

B15, also used red clump giants, and had found that the power spectrum for the $\Delta V_{\rm los}$ component can be minimized for a choice of $V_{\phi,\odot}- V_{c, \odot}=$22.5 \kms{}, i.e., for a value about 10 \kms{} higher than the local standard of rest suggested by \citet{Schonrich:2010}. The peak of the their power spectrum was of the order of this difference of 10 \kms{}, and they attributed this to the amplitude of the streaming motion in the disc. Our estimates of the peak power are still lower by about a factor of 2, compared to B15. In particular we note that the maximum power in the $\Delta V_{\phi}$ component does not exceed 4.2 \kms{}, and so in contrast with B15, we find that the local standard of rest value for the azimuthal Solar peculiar velocity does not require a massive but perhaps a moderate revision. However, given the significant residuals in the radial direction, a revision of that component certainly does not seem unreasonable. 

As demonstration of the high quality astrometry from \gedrthree{}, \citet{klioner2021} used the apparent proper motion of about 1.6 million quasar-like objects, to directly determine that the centripetal acceleration at the Solar System is $\alpha = 5.05 \pm 0.35 \mu$as $yr^{-1}$. Right after, \cite{Bovy:2020} neatly combined this result, with typical values for \Rsun{} and $\Omega_{\rm Sgr}$ (proper motion of Sgr A*), to derive $V_{\odot} = 8.0 \pm 8.4$ \kms{}. While this independent determination of the Solar peculiar velocity is largely in agreement with that by \cite{Schonrich:2010}, the uncertainties on the acceleration are at present too high. Nevertheless, this value would also support that the streaming motion in the azimuthal direction is not very high, as we find in our power spectrum analysis.

\subsection{Connection to Spiral arms}

\begin{table}
\centering
\caption{Pattern speed and Corotation radius for the Milky Way's main Spiral arms. The values for each arm are taken from \citet{Castro-Ginard:2021}, where we have averaged over all ages, and also shown in the brackets, are the values assuming a fixed pattern speed derived by \citet{Dias:2019}. All values are rounded to 1 decimal point for simplicity. \label{tab:spiral_props}}
\begin{tabular}{l|l|l}
\hline
Arm  & $\Omega{_p}$ & CoRotation   \\
     & [\kms{} kpc$^{-1}$]    & [kpc] \\
\hline
\textit{Perseus}  & 19.8 (28.2) &  12.6 (8.9)\\
\textit{Local}  & 32.7 (28.2)&  7.7 (8.9)\\
\textit{Sgt-Car}  & 27.2 (28.2)& 9.2  (8.9)\\
\textit{Sct-Cen}  &  47.7 (28.2)&  5.2 (8.9)\\
\hline
\end{tabular}
\end{table}

In \autoref{fig:cmb_vmap}, on the velocity residual maps, we overplotted the location of the main spiral arms in the Milky Way according to recent models. The concurrence of the profile of a few of the arms and patterns in velocity residuals is interesting to note. For decades there has been a long standing debate over the nature of the Spiral arms in the Galaxy, and very broadly speaking, there are two competing theories. In the standard \textit{density wave theory} proposed by \citet{Lin:1964}, the spiral arms are treated as static density waves that move through the disc with their own pattern speed and are long-lived. The gas and stars in the disc are then slowed down in the arms due to gravitational attraction. Such spiral arms (called \textit{Lin-Shu} type) cannot be
material arms, otherwise they would quickly wind up due to differential rotation and thus break up. On the other hand, is the idea of \textit{Transient} spiral arms. These can be a result of local overdensities that corotate with the disc, and over time shear due to differential rotation. These can also be generated by non static or transient density waves, for example, \citet{D'Onghia:2013} showed that Giant molecular
clouds (GMC) can generate self-perpetuating spiral arms. Understanding the origin of spiral arms is further complicated by the variance observed in their appearance. At both low and high redshift, spiral arms range from the \textit{Grand design} type, where the arms appear highly symmetric and continuous, to the \textit{Flocculent} type, that appear short and fragmented. Their appearance also varies depending on what wavelength and tracers are used to study them. For a more comprehensive review on Spiral arm origin, we point the reader to  \cite{annurevsellwood2022}.

Since we only see the Milky Way edge on we cannot directly observe the spiral arms. However, through a combination of long baseline interferometry (e.g., with the VLBI), accurate parallax and proper motions have been measured for MASERS in high mass star forming
regions, which have broadly been able to map the mutliple arms in the Milky Way, such as the model by \citet{Reid:2019} that we overplot in \autoref{fig:cmb_vmap}. With the availability of high precision 6D phase space for a large number of stars, there have been notable recent efforts in using the dynamics of stars to map the spiral arms in the Milky Way. One of these methods exploits the kinematic and age information of Open Clusters. The basic assumption is that Open Clusters are born in Spiral arms. By integrating backwards, the orbits of present day Open cluster members, the cluster reveals the position of a spiral arm at a past time equal to its age. And by integrating the orbits forward from their birth location, the present day Spiral arm locations can be compared to their analytical predictions. Two recent works have implemented this method, but find completely opposite results. Using \gdrtwo{} data, \citet{Dias:2019} find that the pattern speed of the main spiral arms in the Galaxy are all consistent with each other and have a value around $\Omega_{p}=28.2$ \kms{} kpc$^{-1}$. In contrast, using a much larger, and more recently updated list of Open Clusters, \cite{Castro-Ginard:2021} find that individual arms differ quite a bit in their pattern speeds, and are also able to derive an age-dependence on $\Omega_{p}$. We have compiled the results from the two works in \autoref{tab:spiral_props}, where we also include an estimate for the Co-rotation radius of each arm. According to \citet{Dias:2019}, the Solar neighbourhood is well inside the corotation of all the arms, since they all share the same pattern speed. On the contrary, according to \citet{Castro-Ginard:2021}, only the \textit{Perseus} and \textit{Sgt-Car} arms have their corotation radius beyond the Solar neighbourhood.

\cite{Faure:2014} carried out test particle simulations to predict the global stellar response to spiral perturbations in the Galactic disc, in the absence of an external excitation (such as due to an accreting satellite). They integrate stellar orbits in a 2-arm \textit{Lin-Shu} type spiral potential and produce maps of mean Galactocentric radial velocity (\vR{}). They show (their figure 6.) that inside corotation, in the region traced by the arm, the mean \vR{} is negative (of the order of -7 \kms{}), i.e., stars exhibit bulk motion towards the Galactic center. Meanwhile, in the region between the arms, the stellar radial motion is positive, i.e., stars exhibit bulk motion towards the anticenter. Outside corotation, the pattern is reversed. Their findings confirmed the analytical predictions of \cite{Lin:1964}. In \autoref{fig:cmb_vmap}(b), the local arm clearly overlaps with a region of negative $\Delta V_{R}$. Moreover, adjacent to this is also present, a strong positive $\Delta V_{R}$ feature that also runs from positive to negative $Y_{\rm hc}$. If we assume the fixed pattern speed of the Galactic spiral arms as suggested by \citet{Dias:2019}, this would place corotation at 8.9 kpc, i.e., well beyond the position of the local arm (8.26 kpc). Then the simulations by \cite{Faure:2014} can explain the patterns observed around the Solar neighbourhood. On the other hand, according to the individual arm pattern speeds from \citet{Castro-Ginard:2021}, only the \textit{Perseus}, and the \textit{Sgt-Car} arms have a corotation radius beyond the position of the local arm, and so in that case the \citet{Faure:2014} model does not fit the observed pattern around 8.26 kpc. Furthermore, along the \textit{Perseus} arm, the $\Delta V_{R}$ shows a gradient across the 
$Y_{\rm hc}=0$ line, not seen in the \citet{Faure:2014} maps. The \textit{Perseus} arm also shows a correlation between the radial and the vertical velocity residuals, so the two might be connected in this region. 

A full exploration of the role of the Spiral arms is beyond the scope of this paper, nevertheless, it is interesting to note this overlap of patterns in the velocity residuals and the spiral arm locations, and the diversity of the patterns itself, could be hinting at the different nature of the arms involved. While some of the velocity residuals could be due to spiral perturbations such as those predicted by \cite{Faure:2014}, disrupting/transient spiral arms such as those in our phase-mixing simulation could also have a significant contribution.
Finally, if the principle mechanism responsible for the streaming motion are features such as the Spiral arms of the Galaxy, our results suggest a strong coupling between the in-plane and out-of plane motions \textbf{(such as around the \textit{Perseus} arm)}, and that these extend out to around 1 kpc from the midplane.

\section{Summary and Outlook}

Using a large dataset of about 10 million stars, with 6D phase-space information, we have characterised the velocity field in the Galactic disc out to 3.5 kpc from the Sun. We subtracted axisymmetric models in the individual velocity components, \vphi{},\vR{},\vz{}, and \vlos{}, and performed Fourier analysis to determine the amplitude and physical scale of streaming motion in the disc. We find that the streaming motion is dominant in the \vR{} component, and does not show too much variance with Galactic height $|z|$. The streaming in the \vphi{} component is lower, and drops with \mnz{}. Lastly, the streaming in the \vz{} component, is also lower than in \vR{}, but increases with \mnz{}, likely reflecting the signature of the Galactic warp. The physical scale of the power spectrum implies that the Solar neighbourhood is participating on a large scale streaming motion, and this could have potential implications on the currently assumed Local Standard of Rest. We also find that the predicted location of spiral arms seem to correlate with the patterns observed in the velocity field, and particularly interesting are the correlations between the in-plane and out of plane velocity components. Finally, our test particle simulation of phase-mixing of disrupting spiral arms offers one, of a multitude of physical scenarios that could be causing such streaming motion in the disc.

While our manuscript was under review, the \gaia{} survey had its third data release \cite{gdr3release:2022}, expanding the radial velocity dataset by a factor of four to 32 million sources \footnote{\url{https://www.cosmos.esa.int/web/gaia/dr3}}. Additionally, the survey has also made public Astrophysical parameters for nearly 500 million sources. This truly unprecedented and invaluable chemodynamic dataset is helping piece together the interplay between various structural components and the large scale kinematic processes underway in the Milky Way. We point the reader to \cite{Drimmelgdr3:2022} for a first look at large scale streaming motion in the Galaxy's velocity field with \gdrthree{}.

\section{Acknowledgements}
We thank the anonymous referee for their careful reading of the manuscript, and thought provoking suggestions. 
The authors kindly thank Amina Helmi, Ronald Drimmel, and Jie Yu for their helpful suggestions and comments. SK acknowledges support from the Netherlands Organisation for Scientific Research (NOVA). JBH is supported by an ARC Australian Laureate Fellowship (FL140100278) and the ARC Centre of Excellence for All Sky Astrophysics in 3 Dimensions (ASTRO-3D) through project number CE170100013.

The GALAH survey is based on observations made at the Australian Astronomical Observatory, under programmes A/2013B/13, A/2014A/25, A/2015A/19, A/2017A/18. We acknowledge the traditional owners of the land on which the AAT stands, the Gamilaraay people, and pay our respects to elders past and present. Parts of this research were conducted by the Australian Research Council Centre of Excellence for All Sky Astrophysics in 3 Dimensions (ASTRO 3D), through project number CE170100013.

Funding for Rave has been provided by: the Leibniz Institute for Astrophysics Potsdam (AIP); the Australian Astronomical Observatory; the Australian National University; the Australian Research Council; the French National Research Agency; the German Research Foundation (SPP 1177 and SFB 881); the European Research Council (ERC-StG 240271 Galactica); the Istituto Nazionale di Astrofisica at Padova; The Johns Hopkins University; the National Science Foundation of the USA (AST-0908326); the W. M. Keck foundation; the Macquarie University; the Netherlands Research School for Astronomy; the Natural Sciences and Engineering Research Council of Canada; the Slovenian Research Agency; the Swiss National Science Foundation; the Science \& Technology Facilities Council of the UK; Opticon; Strasbourg Observatory; and the Universities of Basel, Groningen, Heidelberg and Sydney.

Guoshoujing Telescope (the Large Sky Area Multi-Object Fiber Spectroscopic Telescope LAMOST) is a National Major Scientific Project built by the Chinese Academy of Sciences. Funding for the project has been provided by the National Development and Reform Commission. LAMOST is operated and managed by the National Astronomical Observatories, Chinese Academy of Sciences.

Funding for the Sloan Digital Sky Survey IV has been provided by the Alfred P. Sloan Foundation, the U.S. Department of Energy Office of Science, and the Participating Institutions. SDSS acknowledges support and resources from the Center for High-Performance Computing at the University of Utah. The SDSS web site is www.sdss.org.

This work has made use of data from the European Space Agency (ESA) mission
{\it Gaia} (\url{https://www.cosmos.esa.int/gaia}), processed by the {\it Gaia}
Data Processing and Analysis Consortium (DPAC,
\url{https://www.cosmos.esa.int/web/gaia/dpac/consortium}). Funding for the DPAC
has been provided by national institutions, in particular the institutions
participating in the {\it Gaia} Multilateral Agreement.

This research has made use of Astropy, a community-developed core Python package for Astronomy \citep{2018AJ....156..123A}. This research has made use of NumPy (Walt et al., 2011), SciPy, and MatPlotLib (Hunter, 2007).

\section{Data Availability}
The data used in this paper are available upon reasonable request to the corresponding author. All observational data used was obtained from publicly available archives of the individual surveys.




\bibliographystyle{mnras}
\bibliography{biblio.bib} 


\newpage
\appendix

\section{Scattered pixels with high velocity residual in the midplane}
\label{app:scattered_pixels}

\begin{figure*}
\includegraphics[width=2.\columnwidth]{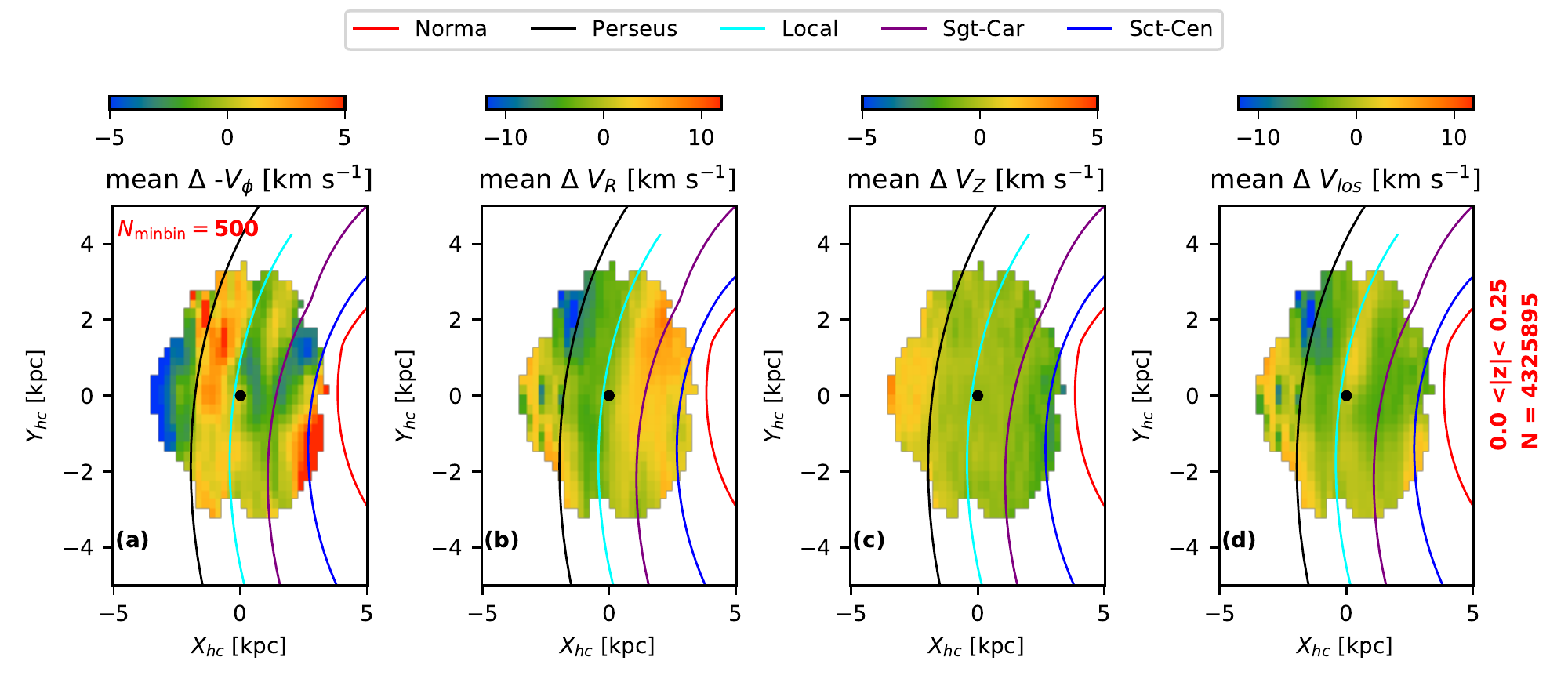} 
\caption{Map of $\Delta V_{i}$ for the \midslice{}, similar to \autoref{fig:cmb_vmap}(a-d), but with $N_{minbin} = 500$. The pixels in  \autoref{fig:cmb_vmap_peculiar}(a-d), are visible even with reduced \textit{Poisson noise}, and are thus statistically significant. \label{fig:cmb_vmap_peculiar}}
\end{figure*}

\section{Best-fit model}
\label{app:best_fit}

Best-fit coefficients with uncertainties for \mnvphi{}$(R,Z)$ as used in equation \ref{eqn:vphi_model}. We neglect the covariance between the coefficients.

\[
  a_{ij}= \qquad \bordermatrix{~  & \tikzmark{harrowleft} a_{00}. & . & a_{02}\tikzmark{harrowright}  \cr
  \tikzmark{varrowtop} a_{00} &-229.15\pm{0.02}  & -0.65\pm{0.05}  & 27.08\pm{0.11}  \cr
  & 0.38\pm{0.02} & -0.49\pm{0.04}  & -8.43\pm{0.10} \cr
  \tikzmark{varrowbottom}a_{20} & 1.09\pm{0.01} & 0.03\pm{0.02} & -0.19\pm{0.05} \cr
                    }
\]
\tikz[overlay,remember picture] {
  \draw[->] ([yshift=3ex]harrowleft) -- ([yshift=3ex]harrowright)
            node[midway,above] {\scriptsize};
  \draw[->] ([yshift=1.5ex,xshift=-2ex]varrowtop) -- ([xshift=-2ex]varrowbottom)
            node[near end,left] {\scriptsize };
}



\bsp	
\label{lastpage}
\end{document}